\documentclass[12pt,preprint]{aastex}
\shorttitle{Relativistic Boltzmann}
\shortauthors{Wolfe}

\begin{document}

\title{The Broadband Spectrum of Galaxy Clusters}

\author{Brandon Wolfe\altaffilmark{1} and Fulvio Melia\altaffilmark{1,2}}
\altaffiltext{1}{Physics Department, The University of Arizona, Tucson, AZ 85721}
\altaffiltext{2}{Steward Observatory, The University of Arizona, Tucson, AZ 85721}

\begin{abstract}
We examine whether nonthermal protons energized during a cluster
merger are simultaneously responsible for the Coma cluster's diffuse radio flux (via
secondary decay) and the departure of its intra-cluster medium (ICM) from a thermal 
profile via Coulomb collisions between the quasithermal electrons and the hadrons. 
Rather than approximating
the influence of nonthermal proton/thermal electron collisions as extremely rare
events which cause an injection of nonthermal, power-law electrons (the `knock-on'
approximation), we self-consistently solve (to our knowledge, for the first time)
the covariant kinetic equations for the two populations. The electron population
resulting from these collisions is out of equilibrium, yet not a power law, and
importantly displays a higher bremsstrahlung radiative efficiency than a pure
power law. Observations with GLAST will test this model directly.
\end{abstract}


\section{Introduction}

Galaxy clusters, aggregates of more than 50 individual galaxies interspersed with
a tenuous plasma known as the intracluster medium (ICM), are the largest gravitationally 
bound objects in the universe. Mergers of such clusters are the most energetic 
astrophysical events since the big bang: a merger of two $10^{15} M_\odot$ 
clusters releases some $10^{63}-10^{64}$ ergs of gravitational energy.

Aside from providing heat (see, e.g., Brunetti et al. 2001; Ohno, Takizawa, \& 
Shibata 2002; Fusco-Femiano et al. 1999; Rephaeli, Gruber, and Blanco 1999) 
observed as bremsstrahlung emission in the ICM (where the X-ray luminosity is 
typically $L_X \sim 10^{45}$ ergs s$^{-1}$, temperatures are in the range 
$\sim 2-10$ keV, and the central density is $\sim 10^{-3}$ cm$^{-3}$) such 
merger events also create supersonic shocks capable of accelerating nonthermal 
particles to energies greater than 1 TeV (Loeb \& Waxman 2000). Examples
of such shocks known to be ongoing in cluster mergers include the galaxy group 
NGC~4839, falling toward the center of the Coma cluster, as observed by XMM-Newton
(Neumann et al. 2001). NGC~4839 achieves a velocity of $\sim 1,400$ km s$^{-1}$, 
and since the sound speed corresponding to Coma's gas temperature of 
$\sim 8$ keV is $\sim 1,000$ km s$^{-1}$, the subcluster's supersonic motion 
is expected to produce shocks, which Neumann et al. (2001) claim to observe 
directly in the imaging of this cluster. Chandra observations of the `bullet' 
cluster 1E~0657-558 (Markevitch 2002, 2004) show a prominent bow shock from 
a lower mass subcluster ($T \sim 6$ keV) as it exits the core of the main 
cluster ($T \sim 14$ keV) at a velocity of $4,500$ km s$^{-1}$. Within such 
shocks, some small fraction---typically  $\sim 5\%$ (see, e.g., Berrington and Dermer 
2005)---of the gravitational energy provided by the merger is believed to
be converted into non-thermal particles through a first order Fermi (Drury 1985) 
process, although neither of the above cases provides direct evidence
for such acceleration.

Further evidence for a population of accelerated, non-thermal particles coexisting
with the thermal background ICM comes from diffuse radio emission extending over
the entire $\sim 1$ Mpc extent of certain rare clusters. Radio luminosities (polarized
in at least one case), $L_r \sim 10^{40}-10^{42}$ ergs s$^{-1}$, have been
measured from emitting regions extending over a Mpc in 30 or so galaxy clusters (see,
e.g., Kim et al. 1990; Giovannini et al. 1993; Giovannini \& Feretti 2000; Kempner
\& Sarazin 2001). This emission---characterized as `halo' if it is radially 
concentrated at the cluster center and `relic' if instead it is located on the 
cluster's outskirts---indicates synchrotron emission of relativistic electrons 
in a magnetized intracluster medium. Radio halos (such as the Coma cluster, which 
takes our focus here) further require a mechanism for constantly replacing energy
lost by nonthermal electrons, since their radiative lifetimes (around $10^8$ years) 
are too short to allow electrons to diffuse across the 100-kpc radio emitting 
region. The radial shape of halo emission, in particular, requires that nonthermal 
electrons be constantly created throughout the cluster.

One natural explanation (Dennison 1980) is that cosmic-ray protons, known to be 
confined within the cluster once produced---perhaps diffused throughout the cluster
following a merger, although a central AGN (Blasi \& Colafrancesco 1999) could also
create sufficient nonthermal hadron energy---constantly collide with hydrogen and 
lose sufficient energy to create pions, which subsequently decay into the 
synchrotron-emitting electrons. However, due in part to difficulties reconciling 
this `secondary' model with all multifrequency observations (Blasi and
Colafrancesco 1999), it is often assumed electrons are constantly reaccelerated 
in situ by a second order Fermi mechanism (see, e.g., Blasi 2000).

In the Coma cluster and a dozen others (Dolang \& Ensslin 2000), the diffuse radio emission exhibits
a spectral break around 1GHz, with a steeper index at higher energy. Entire models have
been based on this break, interpreted as the signature of electron escape (Rephaeli 1979), 
of a reacceleration of already relativistic electrons---possibly energized by a cluster 
merger event (Schlickeiser et al. 1987; Brunetti, Blasi, Cassano, and Gabici 2004)---or 
of energy-dependent radiative losses acting to impose a maximum electron energy
during shock acceleration itself (Berrington \& Dermer 2005; Webb, Drury \& Biermann 1984;
Dolag \& Ensslin 2000). Notably, the secondary model we are proposing here does not reproduce 
this spectral steepening. Instead, it aims to reconcile past X-ray and future $\gamma$-ray
observations using the simplest possible assumptions for the spectrum of accelerated
particles, namely a pure power-law. The model posits that Coulomb collisions between 
the cluster's high-energy proton population, and its thermal electrons, are the root 
of nonthermal excesses in the EUV and X-ray regimes. This is the novel component of 
this picture, replacing the `knock-on' approximation with a full solution to the 
kinetics involved (see \S~2).

Clusters produce an X-ray luminosity $L_X \sim 10^{45}$ ergs s$^{-1}$, generally
interpreted as due to thermal bremsstrahlung by a tenuous thermal plasma characterized
by a temperature in the range $\sim 2-10$ keV and a central density $\sim 10^{-3}$
cm$^{-3}$. These thermal X-ray emitting leptons coexist with the nonthermal
radio-emitting leptons. Several clusters also display
a slight excess above the thermal spectrum starting at $\sim 20-25$ keV and
extending out to energies greater than $\sim 45$ keV. Though still somewhat controversial,
this nonthermal component has thus far been reported for the Coma cluster (Fusco-Femiano
et al. 1999; Rephaeli, Gruber, and Blanco 1999), Abell 2256 (Fusco-Femiano et al. 2000)
and, most recently, for Abell 754 (based on long {\it Beppo}-SAX observations; Fusco-Femiano
et al. 2003). Hard X-ray detections have been reported in several other Abell clusters
(2199, 2319, and 3667) in the redshift range $0.023<z<0.056$, though apparently with
weaker signals. Most of these excesses above the thermal emission can be fitted with
a photon power-law spectrum and index $\sim 2$. A recent example of this fitting
procedure, applied to the RXTE source RX-J0658, may be found in Petrosian (2004).

An interpretation of the X-ray excess (HXR) as inverse Compton scattering (ICS) of 
synchrotron-emitting electrons breaks a degeneracy in the possible values of ICM 
magnetic field and electron energy (see, e.g., Rephaeli 1979). If
the same electrons are responsible for both the radio synchrotron and nonthermal
X-ray emission, then the implied magnetic field $B$ must be an order of magnitude
smaller than that observed. In the Coma cluster, for example, Faraday rotation
measurements suggest that $B\sim 6\,\mu$G (Feretti et al. 1995), in sharp contrast
with the value derived from {\it Beppo}-SAX observations (within the context of the
inverse Compton scattering scenario), which instead require $B\sim 0.16\,\mu$G
(Fusco-Femiano et al. 1999).

It can also be supposed that, in addition to the bremsstrahlung-emitting thermal ICM
and synchrotron-emitting nonthermal electrons, a third nonthermal population of electrons
exists which emits the X-Ray excess (HXR) as bremsstrahlung. Were such a population
thermal, it would require a high temperature ($>50$ keV), impossible to maintain against 
Coulomb re-equilibration if the species were not physically separated. Previously, 
Blasi (2000) suggested that hard X-ray emission in galaxy clusters may be due to 
bremsstrahlung radiation from a population of nonthermal electrons energized 
continuously out of the thermal pool via stochastic acceleration (see also Dogiel 
2000; Sarazin and Kempner 2000; see also Liu, Petrosian, and Melia 2004 for a more
recent treatment of this process). But the re-equilibration timescale of $\sim 1$ Myr, 
in the absence of an efficient energy-loss mechanism, means that all the energy given 
to the nonthermal component has been reprocessed into the thermal pool well before 
$\sim 0.5$ Gyr, heating the ICM above its observed temperature in less than $10^{8}$ 
yrs (Petrosian 2001; Wolfe and Melia 2006b). 

It was, however, suggested by Liang et al. (2002) and Dogiel et al. (2007) that a 
quasi-relativistic third population might overcome this difficulty via a higher 
radiative efficiency (and therefore a longer overheating time). Our model shares
this characteristic, but rather than requiring a second-order Fermi acceleration to 
produce the quasi-relativistic particles, we assume they are produced via collisions 
with nonthermal protons. This appears to be quite natural, since it has long been 
believed that electrons accelerated via collisions with cosmic rays, known as 
`knock-on' electrons, are more abundant than those produced by secondary decay in 
the ($\gamma_e < 100$) region (Schlickeiser 1999; Baring 1991). Our approach is 
simply to solve the kinetic equation of electrons, including both the standard
terms (secondary injection, radiative losses, re-equilibration, etc.) and collisions 
with nonthermal protons. These protons are then simultaneously responsible for both 
the radio-emitting electrons and the nonthermal component of hard X-rays.

An additional motivation for this model is a somewhat controversial excess over 
the expected thermal bremsstrahlung that has been reported in the extreme ultraviolet 
(EUV): in Coma an excess was reported at $1.4 \times 10^{-11}$ ergs cm$^{-2}$ s$^{-1}$ 
(Bowyer et al. 2004; Sarazin \& Lieu 1998). Observations of an EUV excess have also 
been claimed in the Virgo and Abell clusters 1795, 2199, and 4059 (Durret et al. 2002). 
The standard interpretation for these observations is ICS of the cosmic microwave 
background by relativistic electrons. The EUVE measurement is curious because an ICS 
interpretation probes precisely the $10>\gamma_e<100$ portion of the electron 
distribution. Here, we show that secondary electrons alone cannot account for the 
EUVE, because the source function for electrons injected via proton-hydrogen scattering
departs from a pure power-law at the relevant energies. We propose instead that the 
EUVE demonstrates the presence of electrons accelerated to high energies via 
proton-electron collisions.

By far the most significant motivating factor, however, is the possibility of 
observing $\gamma$-rays within a galaxy cluster. While no such observation has 
been confirmed (see, e.g., Reimer 2004), simple arguments (see below) show that 
a positive $\gamma$-ray flux from secondary decays in the Coma and other clusters 
should certainly lie within the sensitivity of the GLAST satellite telescope.

In nine years of observation, the EGRET satellite telescope gave only upper limits 
for the $\gamma$-ray flux of the most X-ray bright galaxy clusters, and it is here 
that our model receives its strongest test. In the case of the Coma cluster 
($<3.8 \times 10^{-8}$ cm$^{-2}$ s$^{-1}$; Reimer 2004), the $\gamma$-ray limit 
{\it excludes} the secondary model from consideration if the magnetic
field within the cluster is $B< 1 \mu$G (Blasi \& Colafrancesco 1999)---i.e., if
the inverse Compton interpretation of the X-ray excess, and not Faraday rotation, 
correctly gives the magnetic field value---because a small magnetic field requires 
a detectably significant population of nonthermal hadrons.

Therefore, the significant contribution of the scenario we present here is the 
reconciliation of the secondary model with high magnetic field values and the 
X-ray excess, in the simplest possible manner. In this picture, cosmic-ray protons 
diffuse evenly throughout the cluster. They collide with hydrogen in the intracluster 
medium, producing a decay cascade whose products include electrons and $\gamma$-rays, 
each peaked at 70 MeV. These electrons, in magnetic fields $> 1 \mu$G, produce diffuse 
radio emission. The cosmic-ray protons also collide with background thermal electrons
and knock the tail of these electrons into a quasi-thermal third population. It is 
this third population, we believe, which is responsible for the EUV and hard X-ray 
excesses.

But this (perhaps still overly simplified) model notably fails to predict the spectral 
steepening at 1 GHz in Coma, as 
well as the radial dependence of the radio spectral index. The competing reacceleration 
model, in which already relativistic relic electrons are re-energized by a second-order 
stochastic acceleration process stirred by cluster mergers, already handles these details 
well. However, a non-ICS interpretation of the X-ray excess will become necessary 
should the GLAST satellite observatory, the HESS telescope---or others---confirm 
a positive $\gamma$-ray flux in the Coma cluster. The model presented here is a 
simple alternative which would be motivated mainly by such a positive detection.

\section{Cosmic-Ray Kinetics}

While our model proposes cosmic ray/electron (cr/e) collisions as a mechanism for producing
the X-ray excess, it is important to stress a distinction between our approach and
the commonly used `knock on' approximation (Abraham, Brunstein, and Cline 1966). 
That model was first proposed to describe an observed population of low-energy 
cosmic-ray electrons which exceeded the number expected from pion decay at the 
70 MeV threshold. To make up for the discrepancy, cosmic ray/electron 
Coulomb collisions were proposed as a dominant mechanism for electron production
in the $20<\gamma_e<200$ regime. 

In the `knock on' approximation, collisions are considered rare events which cause such radical 
change in the electron's energy that electrons simply appear at the higher energy. Such injected
electrons match the proton's spectral index, at a rate proportional to the Coulomb
cross-section for interaction (Baring 1991). Thus to account for the cr/e kinetics, one 
would need to break kinetics in two, with large-angle collisions represented as an 
injection $Q(u)$ and small-angle as the standard Boltzmann equation. More recently, 
however, complete covariant kinetic theories have appeared (Lifshitz \& Pitaevskii 1958;
Nayakshin and Melia 1998; Wolfe and Melia 2006b). If one were to break these theories 
into two parts at the outset---for small and large angles---one would find that the 
large-angle collisions are overwhelmed to the point of irrelevance (Nayakshin \& Melia 
1998), bringing the basic assumption of the Abraham, Brunstein, and Cline (1966) 
approximation into question. And besides, if cr/e collisions were in fact to provide 
a pure power-law injection, they would not be a candidate for producing Coma's hard X-ray 
excess.

Particle collisions between protons and electrons cause the electrons to diffuse to higher
energies, using a formalism very similar to the standard (Rosenbluth et al. 1957) theory.
In that theory, the diffusion and advection of particles in velocity space are given
by convolving the particle distribution with the kernels $|v-v^\prime|$ and $|v-v^\prime|^{-1}$.
In the covariant generalization, these kernels must be corrected to (Landau 1936;
Braams \& Karney 1987; Wolfe \& Melia 2006b)
\begin{equation}
D_{ab}(\mathbf{u}) = \frac{q_a^2 q_b^2}{m_a^2} \Lambda \int Z(u,u^\prime) f(u) du\;,
\end{equation}
\begin{equation}
F_{ab}(\mathbf{u}) = \frac{q_a^2 q_b^2}{m_a m_b}\bigg(1+\frac{m_a}{m_b} \Lambda \int \bigg(
\frac{\partial}{\partial u} Z(u,u^\prime) \bigg) f(u) du\bigg)\;.
\end{equation}
Here and throughout $\mathbf{u} \equiv \gamma \beta c$ is the momentum in units of the rest mass.

Collisions between the nonthermal background of intracluster electrons, and nonthermal hadrons,
are then solved via a coupled Fokker-Planck equation (see Wolfe and Melia 2006b).
It is not difficult to understand how collisions develop the nonthermal electron distribution, 
especially in contrast with the `knock-on' approximation which has, up to now, been used.

Baring (1991) holds that only large-angle collisions provide relativistic boosts to electrons, 
and this forms the basis of the `knock-on' approximations' philosophy. The Fokker-Planck 
equation including knock-on injection is a kind of Frankenstein monster which handles e-p 
collisions once as a flux in the electron's velocity space, and then again as a separate 
electron injection. But knock-on electrons are in fact proposed to describe relatively small 
jumps in energy, up to only $20<\gamma_e<200$, after which secondary production is expected 
to dominate. In writing the covariant Fokker-Planck equation, we have already made a 
small-angle assumption: consider instead Equation (13) in Melia \& Nayakshin (1988) for 
the energy exchange rate before a small-angle assumption is made. As is shown in this
paper, if one were to break this integral into a large and a small angles, the large-angle 
component, which represents `knock on' electrons, would be vastly dominated by energy 
exchanged in small-angle collisions. Proton-electron collisions \emph{do} produce nonthermal 
electrons. But they should be handled within the normal kinetic context, by supposing that 
such collisions require the solution of a coupled Fokker-Planck equation.

Inspection of the diffusion coefficient (Nayakshin and Melia 1998) for a thermal
distribution of electrons and a power-law population of protons reveals that proton collisions
dominate over electron collisions when the electron's momentum is $p \sim m_ec/2$.
This should be compared with the second-order Fermi diffusion coefficient
used in previous studies (e.g., Blasi 2000). Collisional diffusion is less efficient, though of
the same order, in the relevant energy range. Note that, while the power-law diffusion coefficient
$D_{turb}(E) \sim E^\alpha$ will inevitably lead to a power-law tail for the electrons, the 
diffusion coefficient for collisions always falls to a constant, yielding a distribution which 
must go to zero. That is, collisions with a power-law distribution cannot themselves yield 
another power law. Unlike under the `knock-on' approximation, electron collisions with 
power-law protons actually result in a nonthermal electron tail, steeper than the proton 
distribution, and featuring a hard cut-off at high energies. We call such electrons 
`quasi-thermal'.

Distinguishing quasi-thermal electrons from the power-law distributions produced by turbulent 
diffusion or under the traditional `knock-on' approximation is essential, because their 
efficiency of bremsstrahlung emission is different (Liang et al. 2002; Dogiel et al. 2007).

\section{Emission Types}

Our actual calculation is time-dependent and exact. However, we may gain significant grasp of
the model characteristics with a few simple analytic estimates. If we adopt the high-energy
relation of Mannheim and Schlickeiser (1994) for the pion production rate of a power-law
distribution of protons with index $s$,
\begin{equation}
q_{\pi^0}(E_\pi) =
q_{\pi^+}(E_\pi) =
q_{\pi^-}(E_\pi) \sim 13.1\, c\, n_{p0}\, n_{H} \sigma_{pp}
\bigg[ 6(E_\pi/1\;\hbox{GeV}) \bigg]^{-(4/3)(s-1/2)}\;,
\end{equation}
we may easily arrive at an approximate $\gamma$-ray source function,
\begin{equation}
q_\gamma(E_\gamma) = 2 \int^\infty_{ E_\gamma+[m_\pi c^2/(4 E_\gamma)]}
dE_\pi\; q_{\pi^0}(E_\pi) [E_\pi^2 - m_\pi^2 c^4]^{-1/2},
\end{equation}
which we may safely expect (Markoff, Melia, and Sarcevic 1997; Fatuzzo and
Melia 2003; Crocker et al.
2005) will also approximate the number of neutrinos injected per unit
volume, per unit time (both in the high-energy limit). Meanwhile, the rate
of electron injection is
\begin{eqnarray}
q_e&=& \frac{m_\pi}{70\, m_e} q_{\pi^\pm} \bigg(\frac{E_\pi}{70\;\hbox{MeV}}\bigg)
\sim \frac{13}{12}\sigma_{pp}\, c\, n_{H}\, n_{p0}(r) \bigg( \frac{m_p}{24\, m_e} \bigg)^{s_{e0}-1}
(\gamma_e \beta_e)^{-s_{e0}}\;\; {\hbox{cm}}^{-3}\; {\hbox{s}}^{-1}\nonumber \\
            \null&\equiv&K_{inj}\, {\gamma_e}^{-s_{e0}}\;,
\end{eqnarray}
where the electron spectral index $s_{e0} = (4/3)(s-1/2)$ matches that of the pions.
We find that these approximations are valid to within a factor 3 above the threshold energy
for pion production at 70 MeV (Markoff, Melia, and Sarcevic 1997).

Assuming the dominant loss mechanism is either inverse Compton scattering with a blackbody
photon background ($T_{CMB} = 2.73$ K), or radio synchrotron (Wolfe and Melia 2006a),
radiative losses are given
by $-dE_e/dt = a_s E_e^2$, with the constant $a_s = (4/3) \sigma_T\, c\, n_e$ $(\epsilon_{CMB}
+ \epsilon_{B})/m_e c^2$. The energy density ($\epsilon_{CMB}$) in the CMB dominates (by over
a decade) over that ($\epsilon_B$) in the magnetic field.  The equilibrium distribution of
electrons due to injection against these losses is 
\begin{equation}
n(\gamma_e) = \frac{K_{inj}}{m_e c^2 a_s (s_{e0}-1)}\; \gamma_e^{-(s_{e0}+1)} \;.
\end{equation}
The radio synchrotron emissivity (in units of energy per unit volume, per unit time,
per unit frequency) associated with this distribution is then
\begin{equation}
\frac{dE}{dV\, d\nu\, dt} \approx 1.15\, \pi^2 \frac{K_{inj}\, \alpha\, \hbar\,
\nu_B}{m_e c^2\, a_s (se0-1)} \bigg( \frac{\nu_B}{\nu} \bigg)^{s_{e0}/2}\;,
\end{equation}
where $\nu_B$ is the gyrofrequency, and the corresponding Compton scattering emissivity
off the CMB (in units of photon number per unit volume, per unit time, per unit energy) is
\begin{equation}
\frac{dN_\gamma}{dV\, d\epsilon\,dt} = 1.8\frac{r_0^2}{\hbar^3 c^2} \frac{K_{inj}}{m_e c^2 a_s
(s_{e0}-1)} (kT_{CMB})^{(s_{e0}+6)/2} \epsilon^{-(s_{e0}+2)/2}\;,
\end{equation}
where $r_0$ is the classical electron radius, and we have used the fact that the hard
X-radiation is produced below the Klein-Nishina region to simplify the cross section.  Note
that, while the synchrotron emissivity varies roughly as the square of the magnetic field,
inverse Compton scattering depends only on the relative normalization $K_{inj}$ of electrons
required for consistency with radio observations.

Finally, thermal bremsstrahlung emission in the relativistic region is approximated
using a Gaunt factor (Rybicki \& Lightman 1985) representing the multiplicative difference
between quantum-nonrelativistic and QED (Haug 1997) bremsstrahlung cross-sections.
It may be reconstructed to better than $\sim5\%$ accuracy as
\begin{equation}
\bar{g}_{ff}(\phi) = A \log(\phi) + B + C \phi^D + E \phi^F
\end{equation}
where A through B are functions of $\theta$ alone,
\begin{eqnarray}
A,B,E,\& F &=& a_i \theta^{b_i} + c_i \theta^{d_i} + e_i\nonumber\\
D &=& 1.95\nonumber\\
C &=& a_5 \theta^{b_5} + c_5 \theta^{d_5} \exp(-e_5\theta),
\end{eqnarray}
and where $\phi = h\nu/k T$ and $\theta = k T/ m_e c^2$. These coefficients are given in Table 1.
% this is a new parameterization, using 25 rather than 70 numbers and gentler in appearance. And hey,
% you can integrate over temperature in two seconds!

\section{Model Characteristics}

Our guide for reasonable magnetic fields, particle densities, etc. is both a simultaneous
fit of multifrequency observations, and a maximum total energy budget being
the power dissipated in a central
merger event between two clusters of mass $M$, which is
\begin{equation}
L_p \sim 0.1 \frac{G M^2}{R_{sh} t_{cl}} \sim 10^{44}\;\,\hbox{ergs}\;{\hbox{s}}^{-1}\;,
\end{equation}
where the shock radius $R_{sh}$ is $\sim 5$ Mpc, and the cluster's age $t_{cl} \sim 10^{10}$ yr.

Assuming an ICM gas mass $M_{gas}
\sim 10^{14} M_{\sun}$ and a total cluster mass $M_{tot} \sim 10^{15} M_{\sun}$, over an active
distance comparable with the Abell radius, the average ICM gas density is $n_H \sim 3 \times 10^{-4}$
cm$^{-3}$. We adopt a magnetic field $B=0.8\; \mu$G, which would underproduce inverse Compton
scattered photons at the X-ray excess by roughly two orders of magnitude. We accept a recent
estimate for the Coma's distance of 102 Mpc; at a bremsstrahlung temperature $T= 8.21$ keV,
the measured X-ray luminosity then implies an active volume $\sim 1.5$ Mpc$^3$. Finally, we
take the spectral index $s_p$ of the proton distribution $n_p = n_{p0} (\gamma \beta)^{-s_p}$ to
have the value $2.1$. The end result is a nonthermal hadron population
consistent with a cosmic ray luminosity $L_p = 4.5 \times 10^{43}$ ergs s$^{-1}$,
within the energy budget of a cluster merger.

\section{Particle Distribution Function and Broadband Flux}

We show the evolution of the quasi-thermal component of the electron distribution in Fig. 1a,
together with bounding thermal distributions to demonstrate its non-thermal character. The
power-law tail is most pronounced after only $10^{7}-10^8$ years; however, the efficient
energy loss via electron-proton collisions (bremsstrahlung) prevents rapid overheating and
leads to a quasi steady-state distribution. Evolution of the nonthermal component is shown
in Fig 1b: while the high-energy component assumes essentially the form of a pure power-law
injection subject to synchrotron losses, the pion threshold region rapidly thermalizes.

Radio synchrotron emission (Fig. 2a) comes from the purely power-law component of injected
electrons.  Clusters require at least some $100$
Myr to reach their steady state, characterized by a power-law index one greater than the
injection index.  Nonthermal X-ray bremsstrahlung (Fig. 2b) shows an excess consistent with
that observed via {\it Beppo}-SAX (Fusco-Femiano et al. 2004), although less consistent with
the RXTE (Rephaeli \& Gruber 2002) observations. As expected, ICS emission falls well below
the required excess with any magnetic field $\sim 1 \mu$G.

The reported EUV excess (Bowyer et al. 2004; Sarazin \& Lieu 1998) cannot be
due to inverse Compton scattering of CMB photons by the nonthermal component of secondary
electrons, because this emission comes from the threshold region, and where Coulomb
losses dominate. Even supposing a magnetic field of $0.2 \mu$ G, the inverse Compton 
flux at $2\times10^{2}$ eV falls several decades short of the excess power reported by 
Bowyer et al. (2004) (see Fig. 3).

The detection limits of EGRET and GLAST straddle the 70 MeV pion bump, so it is worthwhile
improving upon the rough sketch we have outlined above for the $\gamma$-ray emissivity, by
considering a more exact treatment. In Fig. 4 we display the $\gamma$-ray flux consistent
with both the diffuse radio and X-ray emission. For energies above $E_\gamma > 100$ MeV,
the approximation (thin solid line) holds well. Note that the $\gamma$-ray luminosity
produced by nonthermal electron-proton bremsstrahlung is always several decades below that
produced via pion decay.

\section{Conclusions}

Previously (Liang et al. 2002; Dogiel 2006) it was shown that, because the efficiency
of bremsstrahlung emission for a quasithermal distribution
is decades higher than for a pure power law (Petrosian 2001), this type
of distribution may account for the nonthermal X-ray excess. However, it was believed that
proton collisions did not qualify, since under the `knock-on' approximation they produce electrons
in a power law. Here, by solving the covariant kinetics self-consistently, we have shown
that collisions do create a suitable quasi-thermal distribution.
As further motivation that nonthermal proton/ quasi-thermal electron collisions are in fact
relevant in clusters, we have shown that electrons injected during proton collisions
may not account for the EUV excess via inverse Compton scattering with the CMB,
because the relevant injection occurs below the 70 MeV pion bump.

Describing the X-ray excess as due to a quasi-thermal electron population
allows us to model the Coma cluster of galaxies using a value for the magnetic
field in the intra-cluster medium which agrees with that given by Faraday rotation
measurements. Luckily, our model has a near-term litmus test via the GLAST
satellite: the resolution of $\gamma$-rays from the Coma cluster
would demonstrably prove the presence of nonthermal hadrons and require a secondary
model which could describe radio, X-ray, and $\gamma$-ray observations simultaneously.
A null detection would favor the ICS interpretation of an X-ray excess, meaning that
a secondary model for radio emission in Coma cannot be correct.
Whatever the outcome, we suggest that problems which invoke the `knock-on'
approximation would benefit considerably from a treatment using covariant kinetics,
as we have developed here and in our previous work (Wolfe and Melia 2006b).

\acknowledgements

This research was supported by NSF grant AST-0402502 at the University of Arizona.

\clearpage

\begin{table}[h1]
\caption{Electron-Proton Thermal Gaunt Factor Expansion Coefficients\label{tbl-3}}
\footnotesize
\begin{tabular}{crrrrrrrrrrr}
\tableline\tableline\\
         & $c_{i0}$ & $c_{i1}$ & $c_{i2}$ & $c_{i3}$ & $c_{i4}$ & $c_{i5}$ & $c_{i6}$ \\
\tableline
$c_{0j}$ &9.380e-04      &1.924e-03      &-3.726e-03     &-9.218e-03     &-1.011e-04     &8.037e-03      &4.182e-03 \\
$c_{1j}$ &8.654e-04      &-4.263e-04     &-1.128e-02     &-1.273e-02     &1.150e-02      &2.236e-02      &5.062e-03 \\
$c_{2j}$ &-7.215e-03     &-1.803e-02     &1.731e-02      &6.562e-02      &1.929e-02      &-4.031e-02     &-3.520e-02 \\
$c_{3j}$ &-3.194e-03     &6.941e-03      &6.072e-02      &5.678e-02      &-7.306e-02     &-1.248e-01     &-2.687e-02 \\
$c_{4j}$ &2.083e-02      &5.496e-02      &-4.063e-02     &-1.892e-01     &-7.899e-02     &9.842e-02      &1.279e-01 \\
$c_{5j}$ &-1.647e-03     &-3.669e-02     &-1.110e-01     &-3.922e-02     &2.015e-01      &2.558e-01      &2.765e-02 \\
$c_{6j}$ &-2.376e-02     &-5.335e-02     &8.173e-02      &2.419e-01      &4.259e-02      &-2.123e-01     &-2.705e-01 \\
$c_{7j}$ &1.549e-02      &6.619e-02      &5.372e-02      &-1.291e-01     &-2.818e-01     &-2.152e-01     &1.662e-01 \\
$c_{8j}$ &1.368e-03      &-2.695e-02     &-1.052e-01     &-2.340e-02     &3.103e-01      &5.121e-01      &8.041e-01 \\
$c_{9j}$ &-1.703e-02     &-5.764e-02     &1.883e-02      &2.948e-01      &4.358e-01      &7.627e-02      &2.455e-01 \\
\tableline\tableline\\
\end{tabular}
\end{table}

\clearpage

\begin{figure}
\centering
\plotone{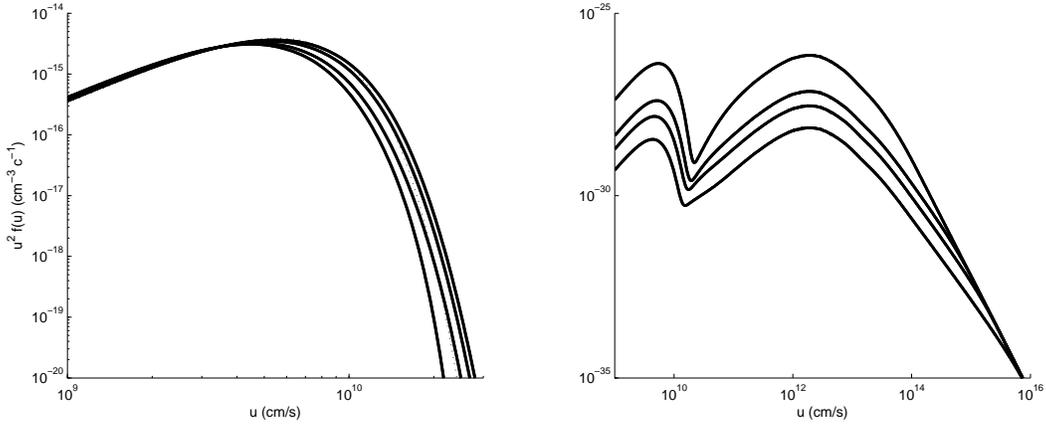}
\caption{
Thermal background electrons (panel a, on the left) at $t = 5$, 20, 102, and 650 Myr (left to
right) are given a nonthermal tail which is most pronounced within 10 Myr, but which is
maintained in a quasi steady-state for upwards of 1 Gyr. A thermal distribution at $T = 8.21$ keV
would lie just inside of the curve at $t=5$ Myr. Nonthermal secondary electrons
(panel b, on the right), at identical times (bottom to top), feature a spectral break due
to ICS losses which by 650 Myr encompasses the energies observed via radio synchrotron.
Electron injection at $E_e < 1$ GeV is limited by the threshold bump, as well as by Coulomb
losses into a thermal population (note the downward curvature near $u = c$ of depleted
electrons by 650 Myr).
}
\end{figure}

\begin{figure}
\centering
\plotone{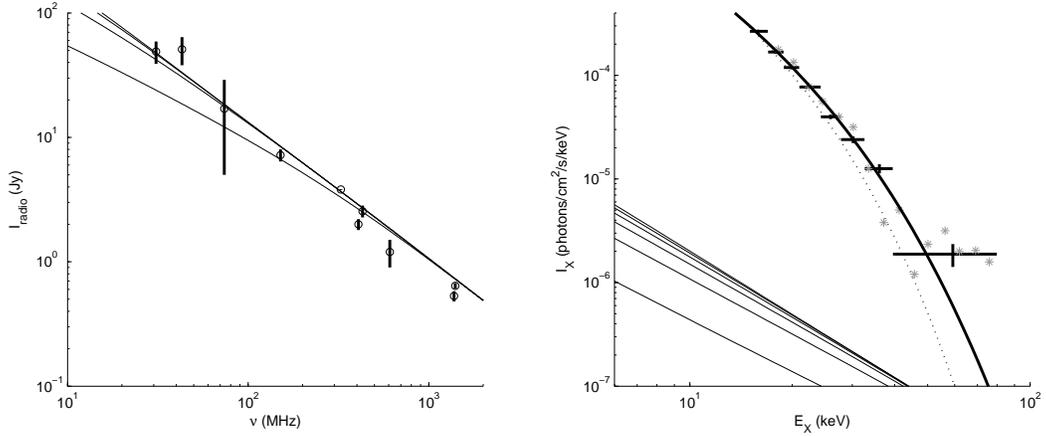}
\caption{
A comparison of the observed diffuse radio emission (panel a, on the left)
below the 1.4 GHz spectral break,
with the calculated synchrotron emissivity at $t = 50$, 100, 150, and 200 Myr, indicates that
the injection must be ongoing for at least 100 Myr. After this time, the distribution has
reached its steady state. Diffuse X-ray emission (panel b, on the right)
from nonthermal electrons at 650 Myr
accounts for the observed hard X-radiation (the histogram is from {\it Beppo}-SAX, the points are
from RXTE)---whereas, at $B = 0.8\;\mu$G, inverse Compton scattered radiation (solid lines to the
bottom left of this panel, shown at times coincident with Fig. 2a) is an order of magnitude below
the hard X-rays. The contribution from a thermal distribution at $T = 8.21$ keV (dotted) is
presented for comparison.
}
\end{figure}

\begin{figure}
\centering
\plotone{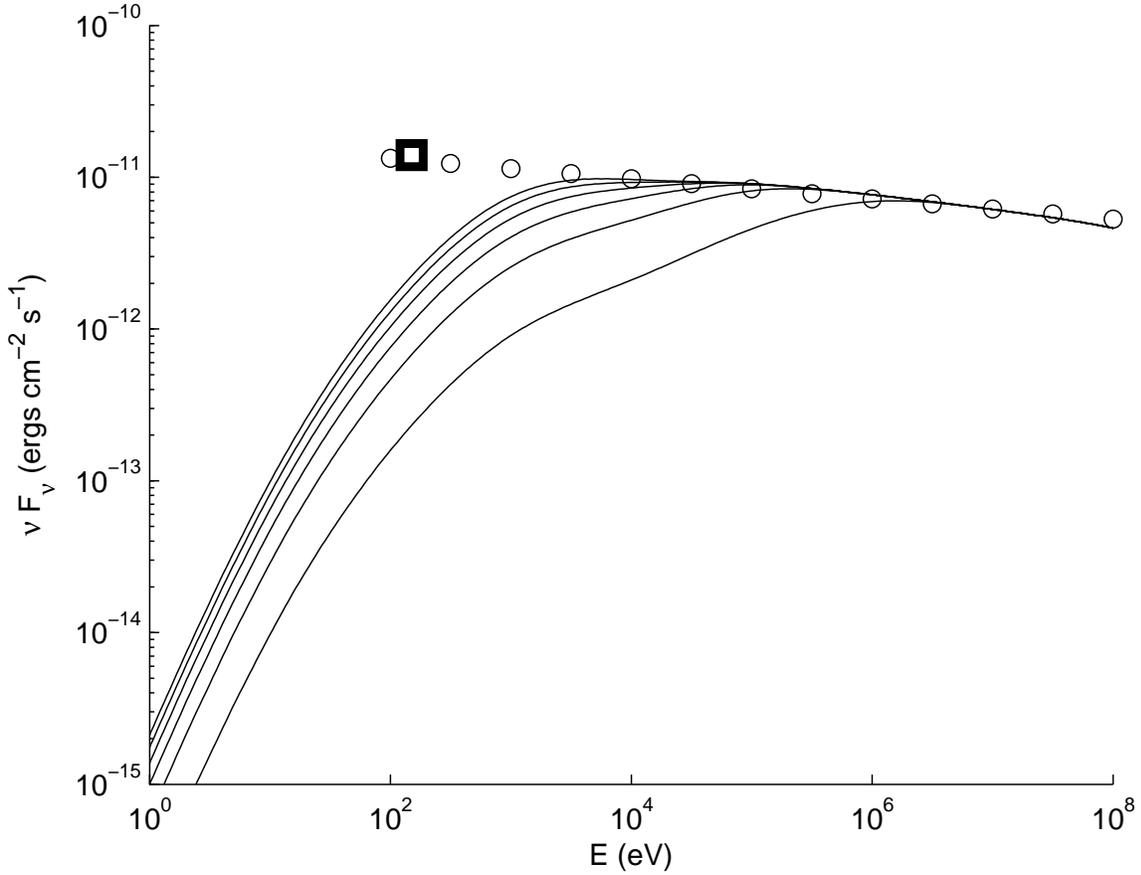}
\caption{
Even supposing a magnetic field of $0.16\;\mu$G, inverse Compton scattering emission from
secondary electrons cannot account for the EUV excess (shown here as a heavy box). The
emissivity due to the self-consistent electron distribution (at t = 50, 150, 250, 350,
450, and 550 Myr) falls below the naive expectation (shown here as circles)
as a result of the pion threshold and Coulomb losses.
}
\end{figure}

\begin{figure}
\centering
\plotone{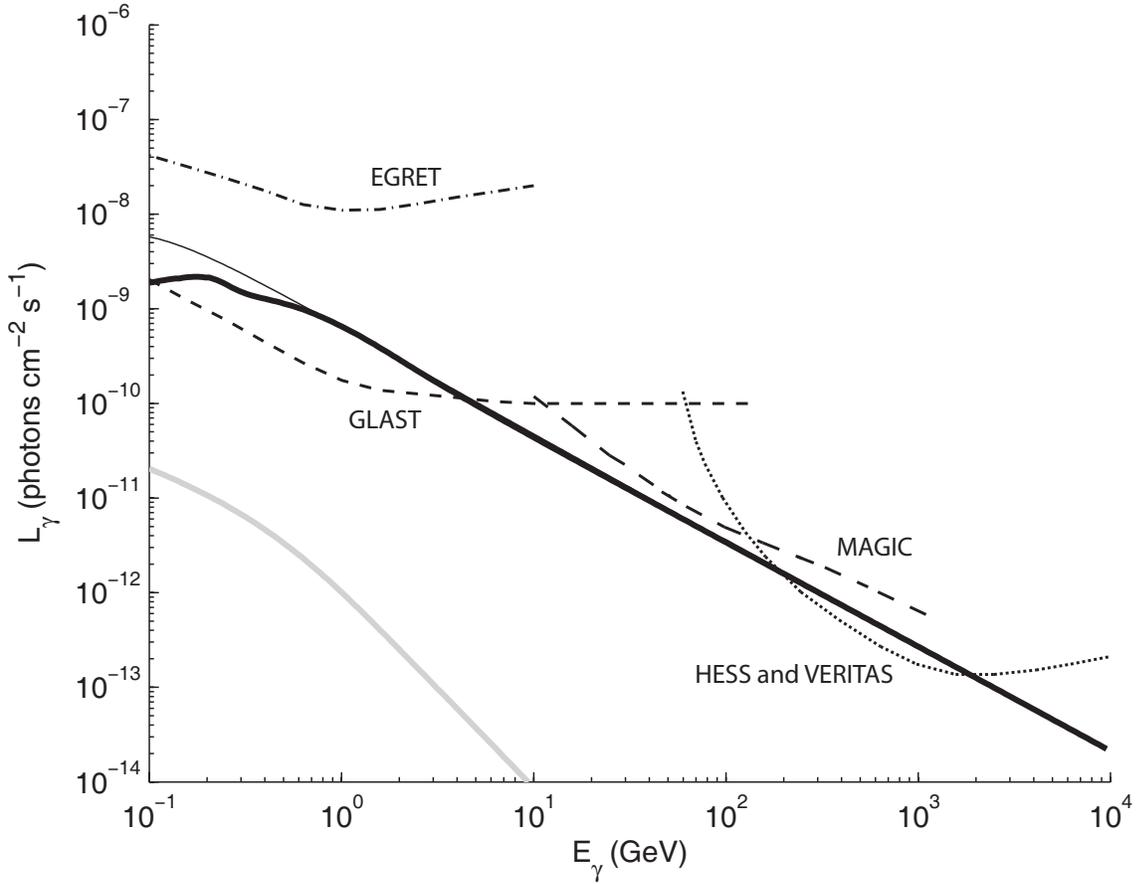}
\caption{
Predicted secondary $\gamma$-ray emission due to proton-proton scattering within
the Coma cluster. Sensitivity limits for EGRET (dot-dash), MAGIC (long dash), GLAST
(short dash), and VERITAS and HESS (dot) (DB05) are also presented. Bremsstrahlung
from the nonthermal electron component (shaded solid) is always well below detection.
}
\end{figure}


\begin{thebibliography}{}
\bibitem[Abraham et al. 1966]{767} Abraham, P. B., Brunstein , K. A., and Cline,
T. L., 1966, Phys. Rev., 150, 1088
\bibitem[Baring 1991]{652} Baring, M., 1991, MNRAS, 253, 388
\bibitem[Berrington and Dermer 2003]{483} Berrington, R. C. and Dermer, C. C., 2003, ApJ,
594, 709
\bibitem[Blasi 2000]{931} Blasi, P., 2000, ApJL, 532, L9
\bibitem[Blasi and Colafrancesco 1999]{485} Blasi, P. and Colafrancesco, S., 1999, Astropart. Phys.,
12, 169
\bibitem[Bowyer et al. 2004]{781} Bowyer, S., Korpela, E. J., Lampton, M., and
Jones, T. W., 2004, ApJ, 605, 168
\bibitem[Braams and Karney 1987]{932} Braams, B. J. and Karney, C. F., 1987, PRL, 59, 1817
\bibitem[Brunetti et al. 2001]{487} Brunetti, G., Setti, G., Feretti, L., and
Giovannini, G., 2001, MNRAS, 320, 365
\bibitem[Brunetti et al. 2004]{489} Brunetti, G., Blasi, P., Cassano, R., and Gabici, S., 2004,
MNRAS, 350, 1174
\bibitem[Crocker et al. 2005]{491} Crocker, R. M. et al., 2005, ApJ 622, 892
\bibitem[Dennison 1980]{492} Dennison, B., 1980, ApJ, 239, L93
\bibitem[Dogiel 2000]{493} Dogiel, V. A., 2000, AA, 357, 66
\bibitem[Dogiel et al. 2007]{595} Dogiel, V. A. et al., 2007, A\&A, 461, 433
\bibitem[Drury 1985]{679} Drury, L. O'D., 1985, in Cosmical gas dynamics, VNU Science Press,
Utrecht, p.~131
\bibitem[Dolag and Ensslin 2000]{999} Dolag, K. and Ensslin, T. A., 2000, A\&A, 362, 151 
\bibitem[Durret et al. 2002]{789} Durret, F., Slezak, E., Lieu, R., Dos Santos, S., 
and Bonamente, M., 2002, A\&A, 390, 397
\bibitem[Fatuzzo and Melia 2003]{494} Fatuzzo, M. and Melia, F., 2003, ApJ, 596, 1035
\bibitem[Feretti et al. 1995]{496} Feretti, L., Dallacasa, D., Giovannini, G., and Tagliani, A.,
1995, AA, 302, 680
\bibitem[Fusco-Femiano et al. 1999]{498} Fusco-Femiano, R., Dal Fiume, D., Feretti, L. et al. 1999, ApJL, 513, L21
\bibitem[Fusco-Femiano et al. 2000]{499} Fusco-Femiano, R., Dal Fiume, D., De Grandi, S. et al. 2000, ApJL, 534, L10
\bibitem[Fusco-Femiano et al. 2003]{500} Fusco-Femiano, R., Orlandini, M., De Grandi, S. et al. 2003, AA, 398, 441
\bibitem[Giovannini et al. 1993]{503} Giovannini, G., Feretti, L., Venturi, T., Kim, K.-T., Kronberg, P. P.,
1993, ApJ, 406, 399
\bibitem[Giovannini \& Feretti 2000]{505} Giovannini, G. and Feretti, L., 2000, New Astronomy, 5, 335
\bibitem[Haug 1997]{597} Haug, E. 1997, A\&A, 326, 417
\bibitem[Kempner \& Sarazin 2001]{506} Kempner, J. C. and Sarazin, C. L. 2001, ApJ, 548, 639
\bibitem[Kim et al. 1990]{943} Kim, K.-T., Kronberg, P. P., Dewdney, P. E., and Landecker, T. L.,
1990, ApJ, 355, 29
\bibitem[Landau 1936]{822} Landau, L. D., 1936, Physik. Zeits. Sowjetunion, 10, 154
\bibitem[Liang et al. 2002]{585} Liang, H., Dogiel, V. A., and Birkinshaw, M., 2002,
MNRAS, 337, 567
\bibitem[Lifshitz \& Pitaevskii 1958]{888} Lifshitz, E. M. and Pitaevskii, L. P.,
1958, Sov. Phys. JETP, 6, 418
\bibitem[Liu et al. 2004]{688} Liu, S., Petrosian, V., and Melia, F., 2004, ApJL,
611, L101
\bibitem[Loeb \& Waxman 2000]{553} Loeb, A. and Waxman, E., 2000, Nature, 405, 156
\bibitem[Mannheim \& Schlickeiser 1994]{665} Mannheim, K. and Schlickeiser, R., 1994,
A\&A, 286, 983
\bibitem[Markevitch et al. 2002]{576} Markevitch, M. et al., 2002  ApJL, 567, L27
\bibitem[Markevitch et al. 2004]{575} Markevitch, M. et al., 2004, ApJ, 606, 819
\bibitem[Markoff, Melia, and Sarcevic 1997]{507} Markoff, S, Melia, F., and Sarcevic, I.,
1997, ApJ, 489, L47
\bibitem[Nayakshin \& Melia 1998]{510} Nayakshin, S. and Melia, F., 1998, ApJS, 114, 269
\bibitem[Neumann et al. 2001]{565} Neumann, D. M. et al., 2001, A\&A, 365, 74 
\bibitem[Ohno et al. 2002]{550} Ohno, H., Takizawa, M., and Shibata, S., 2002, ApJ, 577, 658
\bibitem[Petrosian 2001]{511} Petrosian, V., 2001, ApJ, 557, 560
\bibitem[Petrosian 2004]{512} Petrosian, V., 2004, 35th COSPAR Scientific Assembly, 4202
\bibitem[Reimer 2004]{677} Reimer, O., 2004, Journal of the Korean Astronomical Society,
37, 307
\bibitem[Rephaeli 1979]{513} Rephaeli, Y., 1979, ApJ, 227, 364
\bibitem[Rephaeli et al. 1999]{514} Rephaeli, Y., Gruber, D.E., and Blanco, P., 1999, ApJL, 511, L21
\bibitem[Rosenbluth et al. 1957]{901} Rosenbluth, M.N., MacDonald, W.M., Chuck, W., 1957, 
Phys. Rev., 107, 350
\bibitem[Rybicki and Lightman 1985]{515} Rybicki, G. and Lightman, A., 1985, Radiative Processes
in Astrophysics (Wiley: NY)
\bibitem[Sarazin \& Kempner 2000]{517} Sarazin, C.L. and Kempner, J.C., 2000, ApJ, 533, 73
\bibitem[Sarazin \& Lieu 1998]{791} Sarazin, C. L. and Lieu, R., 1998, ApJL, 494, L177
\bibitem[Schlickeiser 1999]{655} Schlickeiser, R., 1999, A\&A, 351, 382
\bibitem[Schlickeiser et al. 1987]{518} Schlickeiser, R., Sievers, A., and Thiemann, H., 1987, AA, 182, 21
\bibitem[Webb et al. 1984]{731} Webb, G. M., Drury, L. O'C., and Biermann, P., 1984,
A\&A, 137, 185
\bibitem[Wolfe and Melia 2006a]{522} Wolfe, B. and Melia, F., 2006a, ApJ, 637, 313
\bibitem[Wolfe and Melia 2006b]{523} Wolfe, B. and Melia, F., 2006b, ApJ, 638, 125
\end{thebibliography}
\end{document}